\documentclass[aps,prc,amsmath,twocolumn]{revtex4}

\usepackage {bm}
\usepackage{graphicx}

\begin{document}

\title{Nucleon semimagic numbers and low-energy neutron scattering}
\author{D.A.Zaikin\footnote{\normalsize e-mail: zaikin.dmitry@gmail.com} and
I.V.Surkova\footnote{\normalsize e-mail: ivsurkova@list.ru}}
\affiliation{Institute for Nuclear Research of Russian Academy of
Sciences, Moscow, Russia}

\begin{abstract}
It is shown that experimental values of the cross sections of
inelastic low-energy neutron scattering on even-even nuclei
together with the description of these cross sections in the
framework of the coupled channel optical model may be considered
as a reliable method for finding nuclei with a semimagic number
(or numbers) of nucleons. Some examples of the application of this
method are considered.
\end{abstract}

\pacs{24.10.Eq, 25.40.Fq}

\maketitle

\section{Introduction}

During last two decades the problem of existence of so-called
semimagic numbers of nucleons is widely discussed (see {\it e.g.}
\cite{mor}, where one can find a bibliography on this problem).
The existence of semimagic numbers is considered as a result of
the appearance of a new gap in the single-nucleon level scheme by
adding a pair of neutrons (or protons) and its disappearance by
adding another pair. Such an effect is believed to be caused by
the interaction between valence protons and neutrons. For the
first time the surmise about importance of the $np$-interaction
was made by A.De Shalit and M.Goldhaber \cite{DeS} in the shell
model in connection with the appearance of nuclear deformation.
P.Federman and S.Pittel \cite{FP} showed that the interaction
between neutrons and protons with strongly overlapped orbits may
lead to the appearance of nuclear deformation. At the same time
they have shown that this interaction may essentially change the
single-nucleon level scheme. This interaction becomes important
when interacting neutron and proton have big radial or big and
close orbital quantum numbers. Therefore semimagic quantum numbers
are less steady than "classical" ones.

As to methods of finding semimagic quantum numbers of nucleons it
seems that shell model calculations taking into account the
$np$-interaction of valence nucleons can not be regarded as
reliable because the form and intensity of this interaction seem
to be rather uncertain. Therefore semimagic numbers can be found
by indirect ways, namely using a comparative analysis of some
properties of neighboring nuclei, such as biding energies of
nucleons, spectroscopic data ({\it e.g.} {\it g}-factor values,
change of the sign for the coefficient of {\it E2-M1} mixture in
electromagnetic transitions $2^+_2\to 2^+_1$, $3^+_1\to 2^+_1$,
$3^+_1\to 2^+_2$ of even-even nuclei {\it etc}).

In the next section we show that the analysis of inelastic
scattering cross sections of low-energy neutrons on even-even
nuclei gives a reliable method for finding semimagic numbers of
nucleons.

Section 3 presents some examples of the concrete application of
this method. And section 4 contains some conclusive remarks.

\section{Neutron inelastic scattering}

Low-energy neutron data for even-even nuclei with $A\ge 56$ were
successfully  described \cite{ZS99,ZS02} by the authors of this
paper in terms of the coupled channel optical model (CCOM). These
data, taken for neutron energies $E_n\le 3$\,MeV, included total
and elastic scattering cross sections, cross sections of inelastic
scattering corresponding to the excitation of $2^+_1$ level,
angular distributions for elastic and inelastic scattering. We
used two-phonon (five-channel) approximation of CCOM for spherical
nuclei, and three-channel rotational approximation for
nonspherical ones. Details of CCOM are described by
E.S.Konobeevsky, I.V.Surkova  {\it et al} \cite{KS}.

In our calculation we used nonspherical optical potential with the
real part of Woods-Saxon's form. The real part included the
spin-orbital term and the symmetry potential. Radial dependance of
the absorptive part was taken as the derivative of the real part
of the form-factor since namely the nuclear surface is responsible
for absorption of low-energy neutrons. Geometrical parameters of
the potential were the same for real and imaginary parts: the
potential radius was fixed as $R=r_0A^{1/3}$ with $r_0=1.22$\,fm,
the nuclear diffuseness parameter was initially chosen to be equal
to 0.65\,fm, but for some nuclei (see below) it was somewhat
changed. The spin-orbit interaction parameter $V_{so}$ was equal
to 8\,MeV. The real part of our potential included the isotopical
term proportional to $(N-Z)/A$, so that the depth of the real part
of the potential had the following form:

\begin{equation}
 \label{a}
V=V_0-V_1\frac{N-Z}{A},
\end{equation}
where $V_1=22$\,MeV. The depths $V_0$ (of real part) and $W$ (of
imaginary part) were two free parameters to fit. The values of the
quadrupole deformation parameter $\beta_2$ were taken from
compillation \cite{Ram}.

The optimal description of all the totality of low-energy neutron
data for even-even nuclei under consideration was achieved using
practically the same values of model parameters $V_0$ and $W$,
namely, $V_0=52.5\pm 1.5$\,MeV and $W=2.5\pm 0.5$\,MeV. As to
diffuseness parameter $a$, in order to get a good description of
all the nuclei under consideration (including nonspherical ones)
with the same values of $V_0$ and $W$, we had to change the
diffuseness parameter for nonspherical nuclei from the initial
value of 0.65\,fm up to 0.70 --- 0.75\,fm. For the same reason we
slightly diminished the value of $a$ for some spherical nuclei
which were found to be magic or semimagic, taking $a=$0.55 --
0.60\,fm for magic and semimagic nuclei and $a\approx 0.50$\,fm
for double-magic ones.

These alterations in numerical values of $a$ seem to be consistent
(at least, qualitatively) with the experimental data and
theoretical view on the thickness of nuclear surface lay; on the
other hand, they are important for finding semimagic numbers of
nucleons (see below).

Note, that using CCOM with these parameter values we also obtained
a good description \cite{ZS99,ZS02} of the experimental data on
$s$-, $p$- and $d$-neutron strength functions and potential
scattering lengths for nuclei under consideration.

The analysis of inelastic neutron cross sections led us to
so-called $N_{\rm p}N_{\rm n}$-systematics \cite{ZS99,ZS94}: we
showed that the inelastic cross section with the $2^+_1$ level
excitation, taken at the energy equal to 300\,keV over the
threshold and averaged over the energy range of 100\,keV, seems to
be a smooth function of $N_{\rm p}N_{\rm n}$ --- the product of
valence proton and neutron numbers (or their holes). The curve,
presenting this function $\sigma_{\rm inel}(N_{\rm p}N_{\rm n})$
and shown with experimental points in those publications, was
obtained using the least square method. The corrected and more
precise version of this plot is given by fig.1 of this paper.
Note, that the curve presenting a function $\sigma_{\rm
inel}(N_{\rm p}N_{\rm n})$ may also be well described as
\begin{equation}
\label{b} \sigma_{\rm inel}=A+B\sqrt{N_{\rm p}N_{\rm n}}
\end{equation}
with $A=$0.5\,b and $B=$0.14\,b. Emphasize, that this
approximation is not good for big ($>120$) values of $N_{\rm
p}N_{\rm n}$.

\begin{figure}
\includegraphics[width=8cm]{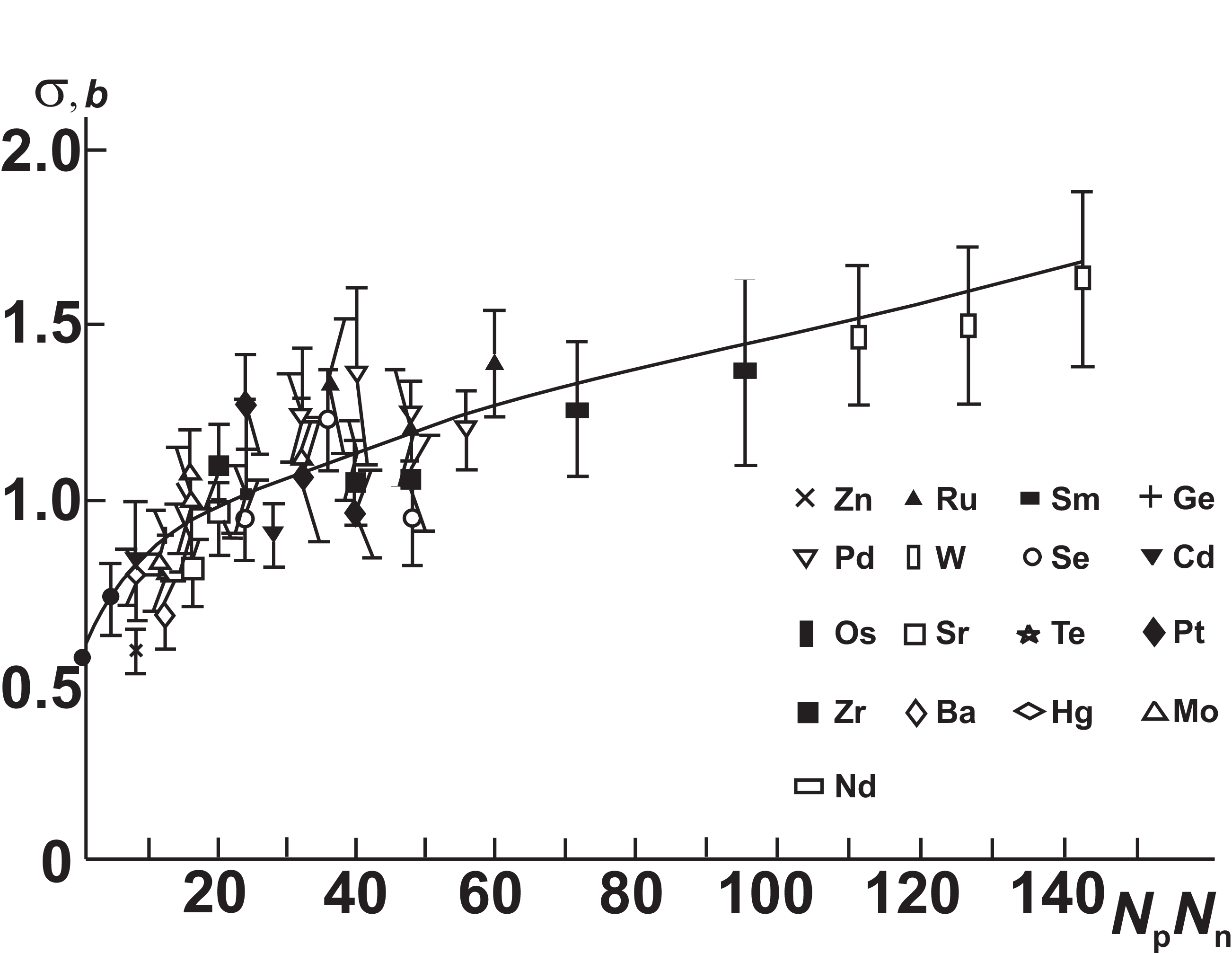} %
\caption{Neutron inelastic cross sections for $E_{\rm n}=300$\,keV
over threshold {\it vs.} the $N_{\rm p}N_{\rm n}$-product}
\end{figure}

Since the $N_{\rm p}N_{\rm n}$-product depends on the beginning
and the end of the upper shells of the nucleus, the $N_{\rm
p}N_{\rm n}$-systematics combined with CCOM-calculations may serve
as a method of finding semimagic numbers of nucleons. The main
scheme of this method is as follows.

If the cross section of neutron inelastic scattering for an
even-even nucleus is essentially less than for neighboring
even-even nuclei (usually it is near by 0.5\,b) and therefore the
corresponding experimental point on the plot of fig.1 deviates
from the curve $\sigma_{\rm inel}(N_{\rm p}N_{\rm n})$, we change
the beginning (or the end) of valence shell in order "to restore
the agreement". As a result of this operation, we obtain a new
value of the valence nucleon product and thereby
--- a new number of nucleons corresponding to the beginning (or
the end) of valence shell, {\it i.e.} a new magic number. This
result may be considered as preliminary one; to confirm it we can
see what is the result of the CCOM-description of this nucleus,
namely, what is the value of diffuseness parameter: if the value
of the diffuseness parameter is essentially less than 0.65\,fm
(0.50
--- 0.60\,fm) it means that the preliminary result is confirmed
and a new semimagic number is found, if not --- it is necessary to
use another method to confirm or to reject the preliminary result.
Such an important role of the diffuseness parameter in our method
is based on the high sensitivity of the CCOM-calculation results
to the diffusion parameter value.

Thus, our method is expected to be efficient if neutron inelastic
cross sections for the considered nucleus and for its "neighbors"
are known with an appropriate accuracy. As we will show, in
absence of such data it is also possible to obtain some results,
but, may be, not so certain as in the main scheme of the method.

Next section is devoted to practical applications of the method
proposed here.

\section{Semimagic numbers of nucleons}

Here, using the method described in the previous section, we
prove the existence of neutron semimagic numbers $N_{\rm s}= $38,
56, 64, and proton semimagic numbers $Z_{\rm s}=$ 40, 58, 64. We
also consider possibilities of existence of some other semimagic
numbers.

\subsection{N$_{\rm s}$=38 and Z$_{\rm s}$=40}

The case of $N_{\rm s}=38$ seems to be the best illustration of
our method application \cite{ZS99,ZS02} (see also \cite{ZS05}).
The neutron inelastic cross sections for neighboring isotopes
$^{70}$Ge and $^{72}$Ge differ by factor 1.7. So, following our
scheme we have to assume that the $N_{\rm p}N_{\rm n}$-values for
these isotopes are equal 0 and 40 respectively. But it means that
in $^{70}$Ge the neutron state $p_{1/2}$ is shifted up, so that a
considerable energy gap appears between it and low-lying state
$f_{5/2}$; this gap disappears by adding a pair of neutrons, {\it
i.e.} in $^{72}$Ge. In other words, $N_{\rm s}$=38 is a semimagic
number. Calculation of diffuseness parameter $a$ for $^{70}$Ge
confirms this conclusion (the value $a=0,65$ presented in
\cite{ZS99} does not take into account "the correction" of $V_0$
and $W$; taking into account this effect gives $a=0.58$).

As to $Z=40$ it seems that in distinction from $N=38$ the proton
state $p_{1/2}$ is not shifted up, and the energy gap appears
between this state and the state $g_{9/2}$ (which is higher than
$p_{1/2}$). By adding a new pair of proton this gap disappears as
a result of the interaction between valence protons and neutrons
in states $g_{9/2}$ and $g_{7/2}$. Thus, for protons we obtain
$Z_{\rm s}$=40. The standard (for our method) analysis of the
low-energy neutron interaction with isotopes of Sr, Zr and Mo
confirms this conclusion.

Note, that the conclusion about $N_{\rm s}$ and $Z_{\rm s}$
follows immediately from the investigation by P.Federman and
S.Pittel \cite{FP}. Later on many authors presented arguments in
favor of {\it a possibility of existence} of semimagic numbers
$N=38$ and $Z=40$ (see {\it e.g.} \cite{Buc}). But in distinction
from those arguments our method gives evidence for {\it the
existence} of these semimagic numbers.

\subsection{N$_{\rm s}$=56 and Z$_{\rm s}$=58}

To some extent the situation with semimagic numbers 56 and 58 is
similar to the case of numbers 38 and 40. The standard analysis of
the experimental data on the interaction of low-energy neutrons
with even-even isotopes of Sr, Zr and Mo \cite{ZS05} leads to the
conclusion that for the isotopes with 56 neutrons the energy level
of the neutron state $2d_{5/2}$ is shifted down and the level of
$1g_{7/2}$-state is shifted up, so that a gap appears between
these states and disappears by adding two more neutrons. In other
words, $N=56$ is a semimagic number. One of the first conclusion
about a possibility of the existence of $N_{\rm s}=56$ was made by
V.A.Morozov \cite{mor}, who showed that at $N=56$ the coefficient
of mixture of $E2$ and $M1$ in electromagnetic transitions
$2^+_2\to 2^+_1$, $3^+_1\to 2^+_1$, $3^+_1\to 2^+_2$ changes the
sign. And since such a change normally is treated as the presence
of a filled subshell, he assumed that the existence of a semimagic
number $N=56$ is quite possible. Our result (being more definite)
confirmed this assumption.

In this connection we should mention recent works by Moscow
University group \cite{Bob} in which it was in particularly shown
that the energy gap between neutron subshells in $^{96}$Zr may
achieve $\sim 3$\,MeV. As a result the authors assumed {\it a
possibility of existence} of the semimagic neutron number $N_{\rm
s}=56$. This calculation may be regarded as an additional
confirmation for our result.

As to a semimagic number of protons $Z_{\rm s}$=58 or 56 some
authors (see {\it e.g.} [1]) assumed that such a number may exist.
This assumption was confirmed in the framework of our approach.
Trying to describe neutron data for Nd we discovered that neutron
inelastic cross sections for some isotopes do not enter the
$N_{\rm p}N_{\rm n}$-systematics. For instance, from the point of
view of this systematics the $N_{\rm p}N_{\rm n}$-value for the
isotope $^{146}$Nd must be not equal to 40 (if its proton shell
would start at $Z$=51), but 8 or 4, and these values correspond to
$Z_{\rm s}$=58 or 56 (see fig.2).

\begin{figure}
\includegraphics [width=8cm]{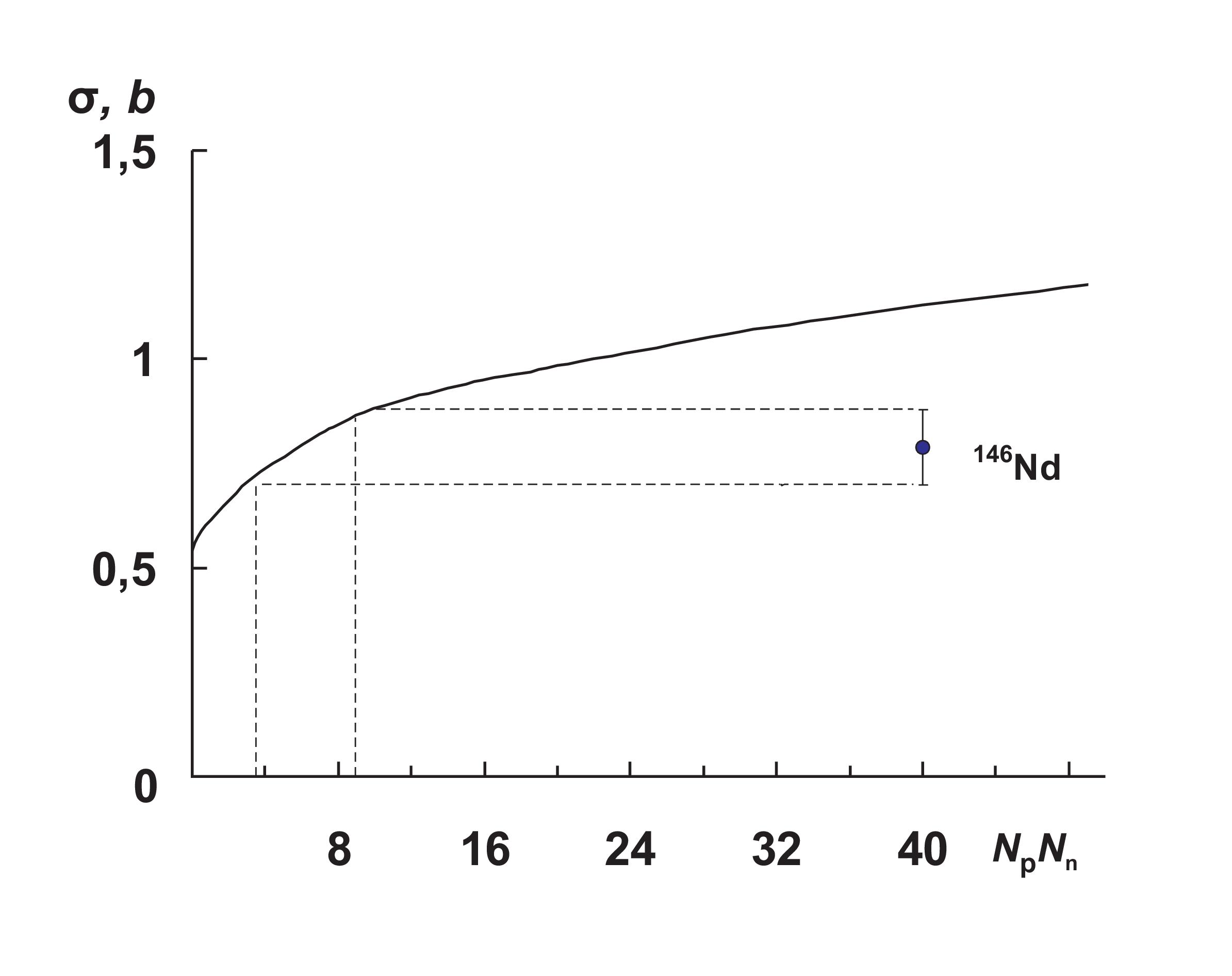}
\caption{Description of neutron date for $^{146}$Nd requires the
$N_{\rm p}N_{\rm n}$-product to be equal not to 40, but to 4 or 8
which correspond to existence of semimagic numbers $Z$=58 or 56}
\end{figure}

Considering isotopes of Ce ($Z$=58) and Ba ($Z$=56) shows that
inelastic cross sections at the energy 300\,keV above the
threshold are somewhat lower for isotopes of Ce than for isotopes
of Ba. For instance, for $^{140}$Ce this cross section is equal to
0.60\,b but for $^{136}$Ba --- 0.66\,b. Diffuseness parameter $a$
calculated by use of CCOM for isotopes of Ce is equal to 0.57 --
0.60\,fm, but for isotopes of Ba --- somewhat bigger (0.60 --
0.65\,fm).

Thus, our results confirm the assumption about existence of the
semimagic number $Z_{\rm s}$=58(56). At the same time they show
that $Z_{\rm s}$=58 is more preferable than $Z_{\rm s}$=56.

\subsection{N$_{\rm s}$=Z$_{\rm s}$=64}

In distinction from the semimagic numbers considered above, the
assumption about existence of the semimagic number 64 leads to the
conclusion that some nuclei (namely $^{114}$Sn and $^{146}$Gd)
must be double magic.

In 1953 J.O.Rasmussen with colaborators \cite{Ras} payed attention
to anomalies in energies of $\alpha$-decay of nuclei with $Z$=64
at $N\approx$ 82 and thus assumed that $Z$ =64 is a semimagic
number. Later on it was assumed \cite{Cas,Ciz} that the subshell
$Z$=64 exists in the region of rearearth nuclei and disappears in
consequence of the $pn$-interaction between $h_{11/2}$- and
$h_{9/2}$-states. Many authors (see {\it e.g.}
\cite{mor,ZS94,Cas,ZS98,Isak}), considering the question about
$Z_{\rm s}$, brought different arguments in favor of existence of
such a semimagical number. In particular, the description of the
neutron inelastic scattering for the isotopes of Sm in the
framework of CCOM is possible only under assumption of the
disappearance of the shell ($1g_{7/2}$ and $2d_{5/2}$) gap at
$Z=64$ and $N>80$ \cite{ZS94}. At the same time the value of
diffuseness parameter $a$ at the optimal description of the
neutron cross sections for $^{146}$Gd is equal to 0.50\,fm, while
for neighboring nuclei $a$=0.55 -- 0.60\,fm. Such values of $a$
give an evidence that $^{146}$Gd is a double magic nucleus, {\it
i.e.} $Z$=64 is a semimagic number.

The situation with $N_{\rm s}$=64 is similar to that with $Z_{\rm
s}$=64. Many authors  \cite{mor,ZS99,Cas,ZS98} concluded that the
existence of this semimagic number is quite possible. The analysis
of neutron inelastic scattering on the isotopes of Cd in the
framework of the $N_{\rm p}N_{\rm n}$-systematics confirmed that
conclusion (just like such an analysis for Sm confirmed the
existence of $Z_{\rm s}$=64). The value of the diffusion parameter
for $^{114}$Sn ($a$=0.50\,fm) gives an argument in favor of the
conclusion that $^{114}$Sn is a double magic nucleus (just like
$^{146}$Gd).

Thus, we may conclude that the existence of semimagic numbers
$Z_{\rm s}=N_{\rm s}=64$ is firmly proved.

\subsection{Other possible semimagic numbers}

Besides the semimagic numbers of nucleons considered above there
are some "candidates" for joining the "semimagic community".
Unfortunately, the method described here cannot give so far any
certain answer about their existence because in most cases the
accuracy of necessary experimental data is insufficient.
Nevertheless, the use of it can lead to some conclusions (at
least, on the level of assumption).

As the first example of such a "candidate" we consider $Z_{\rm
s}$=76 \cite{mac}. Values of the neutron inelastic cross section
for isotopes of Os ($Z=76$) are essentially bigger than 0,5\,b and
the diffuseness parameter values (corresponding to the best
description of low-energy neutron data) are equal to 0.65. All
that testifies against the existence of this semimagic number.

 It seems to be interesting to investigate the region of heavy
($Z>92$) nuclei in connection with existence of semimagic nucleon
numbers. But since for such nuclei the inelastic cross sections of
low-energy neutron scattering are not known, it is impossible to
use the main scheme of our method. However, one can try to draw
some plausible assumptions (or at least some hints) from available
data on inelastic neutron scattering on $^{232}$Th. As a matter of
fact there exist two sets of the experimental data on neutron
inelastic scattering for this nucleus. These cross section values
corresponding to $2^+_1$ level excitation for the neutron energy
300\,keV over the threshold (after averaging over the energy range
of 100\,keV) are equal to 1.13$\pm$0.21\,b and 1.22$\pm$0.22\,b
\cite{mant}. Both of these values (corresponding to $N_pN_n=128$)
are essentially less than the value required by the
$N_nN_n$--systematics. If we (ignoring the experimental errors)
assume that the true value of the cross section considered here is
situated between 1.13 and 1.22\,b, we obtain that the
$N_pN_n$-product must be equal to 48 (instead of 128). Such a
change may be correct only if some new semimagic numbers exist. We
will try to find them.

Generally speaking, the value of the $N_{\rm p}N_{\rm n}$-product
equal to 48 may leads to 6 possible pairs of valence nucleon
numbers $N_p$ and $N_n$. But it is necessary that between the new
and the old values of these number the following relations were
fulfilled:
\begin{equation}
\label{c} N^{\rm new}_{\rm p}\le N^{\rm old}_{\rm p},\quad N^{\rm
new}_{\rm n}\le N^{\rm old}_{\rm n}.
\end{equation}
The breach of one of these requirements would lead to drastic
changes in the nucleon level schemes, in particular, to the
removal of classical magic numbers. (In principle, such a
reconstruction of the nucleon scheme is possible. But there is no
reason for it so far.). Therefore only three pairs $(N_{\rm
p},\,N_{\rm n})_{\rm k}$ (where $k=1,\,2,\,3$), namely, (8,\,6),
(6,\,8), (4,\,12) have to be taken into consideration. For each
pair it is easy to get the nucleon numbers $Z_{\rm k}$ and $N_{\rm
k}$ from which one has to count the values of relevant valence
nucleon numbers:
\begin{equation}
\label{d} \left.
\begin{array}{c}
{\displaystyle Z_1\choose \displaystyle N_1}={\displaystyle
{82\,{\rm or}\,98}\choose \displaystyle {136\,{\rm
or}\,148}},\\
{\displaystyle Z_2\choose \displaystyle N_2}={\displaystyle
{84\,{\rm or}\,96}\choose \displaystyle {134\,{\rm
or}\,150}},\\
{\displaystyle Z_3\choose \displaystyle N_3}={\displaystyle
{86\,{\rm or}\,94}\choose \displaystyle {130\,{\rm or}\,154}}.
\end{array}
\right\}
\end{equation}

Eqs.(\ref{d}) contain 12 numbers $Z_{\rm k}$ and $N_{\rm k}$.
Using these numbers one can form 12 pairs ($Z_{\rm k}$, $N_{\rm
k}$), and if both numbers of a pair are semimagic ones a necessary
value of the $N_{\rm p}N_{\rm n}$-product (in our case it is 48)
will be guaranteed. It remains to prove that at least there is one
such a pair among the pairs considered here. In this connection
note that one of the $Z_1$-values is equal to 82 which is a
classical magic number, so that in this case it is only necessary
to prove that one of 136 or 148 is a neutron semimagic number.

Unfortunately, it seems impossible to find a "semimagic pair"
among those 12 ones for lack of necessary experimental data.
Nevertheless, we shall try to draw some conclusions from available
data (at least on the level of assumptions). Let us consider 12
nuclei consist of $Z_{\rm k}$ protons and $N_{\rm k}$ neutrons
($Z_{\rm k}$ and $N_{\rm k}$ belong to the same pair). 10 of these
nuclei are not interesting for us because some of them are exotic
ones (like $^{228}$Pb) and for others nothing is known about their
interaction with low-energy neutrons. Remaining  two nuclei,
namely $^{246}_{98}$Cf$_{148}$ and $^{246}_{96}$Cm$_{150}$ could
be interesting from the point of view of our method. But we do not
know anything about interaction of the isotopes of Cf with
low-energy neutrons. So, only $^{246}$Cm remains to be considered.
For this and some other isotopes of Cm the neutron strength
functions are known experimentally. The strength function $S_0$ of
$s$-neutrons for $^{246}$Cm is equal to 0.45$\pm $0.15 while for
neighboring isotopes $^{244}$Cm and $^{248}$Cm $S_0$=1.00$\pm
$0.20 and 1.10$\pm$0.12 \cite{ref}. Such a big difference between
the $S_0$-values for $^{246}$Cm and its neighboring even-even
isotopes may be considered as an argument in favor of the
assumption that $N$=150 is a semimagic number. The difference
between the $S_0$-values for $^{246}$Cm and $^{244}$Pu is of the
same order of magnitude, but we cannot say anything about the
other "neighbor on the $Z$-line", {\it i.e.} $^{248}$Cf because
its interaction with low-energy neutrons is not known. So, we can
only say that such a situation may be considered not as an
argument, but only as a "hint in favor" of the assumption that
$Z$=96 is also a semimagic number.

Generally speaking, almost all the $S_0$-values in the
(Pu--Cm)-region are equal to $\approx $1 \cite{ref}. The only
exception is $^{246}$Cm ($S_0=0.45\pm 0.15$). Such a situation
provokes us to assume that $^{246}$Cm is a double magic (or double
semimagic) nucleus. In other words, both $Z=96$ and $N=150$ are
semimagic numbers. Emphasize, that this is only a statement about
the possibility of existence of two new semimagic numbers and
needs more rigorous proof. Unfortunately, the diffusion parameter
value taken from our calculation of $S_0$ cannot help because such
calculations are less sensitive to the diffusion parameter value
than calculations of cross sections.

\section{Conclusion}

The efficiency of the method for finding new semimagic numbers of
nucleon, based on the analysis of low-energy neutron data, was
demonstrated here by different examples of its application. Using
this method we obtained some already known results about existence
of some semimagic numbers, confirmed assumptions about other
semimags, for the first time proved the existence of the semimagic
number $Z_{\rm s}$=58. Using scarce experimental data we first
pointed on the possibility of existence of semimagic $Z=96$ and
$N=150$. Thus, this method seems to be rather efficient.

It is necessary to emphasize that the most important part of the
method described here is the $N_{\rm p}N_{\rm n}$-systematics,
{\it i.e.}, smooth dependence of neutron inelastic cross section
on the $N_{\rm p}N_{\rm n}$-product. This dependence itself is out
of any doubt: there is no one serious contradiction with it.
Therefore this method  becomes less efficient in cases when one
cannot address the $N_{\rm p}N_{\rm n}$-systematics, {\it i.e.}
when neutron inelastic cross sections are not known or their
accuracy is poor.

Also we mention another essential element of our method having
auxiliary character, namely the diffusion parameter value obtained
from the requirement of optimal description of neutron data in
framework of the coupled channel optical model. But this auxiliary
element is efficient only if corresponding calculations are
sensitive to this value. For instance, it works well for the cross
section calculations, but does not work for calculations of
strength functions (as it was mentioned above). Nevertheless, even
in such cases our method permits to obtain some results.

 Thus, the method described here may be considered as one of the
most efficient methods for finding new semimagic (or magic) number
of nucleons.
 \centerline{\bf Acknowledgement}

The authors are indebt to M.V.Mordovskoy for preparing graphic
materials.

\end{document}